\documentclass[notoc]{JHEP}
\usepackage{psfig}
\usepackage{bm}
\usepackage{graphicx}

\newcommand{\eps}  {\varepsilon}

\preprint{hep-ph/0508299, MPP-2005-85}

\title{Probing new physics with coherent neutrino scattering off
nuclei}

\author{J. Barranco \\
Departamento de F\'{\i}sica, Centro de Investigaci{\'o}n y de 
  Estudios Avanzados del IPN, Apdo. Postal 14-740 07000 
  M\'exico, D F, M\'exico\\
\email{jbarranc@fis.cinvestav.mx}}
\author{O. G. Miranda\\
Departamento de F\'{\i}sica, Centro de Investigaci{\'o}n y de 
  Estudios Avanzados del IPN, Apdo. Postal 14-740 07000 
  M\'exico, D F, M\'exico\\
\email{Omar.Miranda@fis.cinvestav.mx}}

\author{T. I. Rashba \\
Max-Planck-Institut f\"ur
Physik (Werner-Heisenberg-Institut), F\"ohringer Ring 6, 80805
M\"unchen, Germany\\
and Institute of Terrestrial Magnetism,
  Ionosphere and Radio Wave Propagation of the Russian Academy of Sciences,
  142190, Troitsk, Moscow region, Russia
\email{timur@mppmu.mpg.de}}

\abstract{
The possibility off measuring for the first time neutrino-nuclei
coherent scattering has been recently discussed by several
experimental collaborations. It is shown that such a measurement may 
be very sensitive to non-standard interactions of neutrinos with
quarks and might set better constraints than those coming from
future neutrino factory experiments. We also comment on other
types of new physics tests, such as extra heavy neutral gauge bosons,
where the sensitivity to some models is slightly better than the
Tevatron constraint and, therefore, could give complementary
bounds. }

\keywords{Neutrino Physics  Beyond Standard Model  
Neutrino Detectors and Telescopes}
\begin{document}

\section{Introduction}

Coherent neutrino-nucleus~\cite{Freedman:1973yd} and
neutrino-atom~\cite{Gaponov:1977gr,Sehgal:1986gn} scattering have been
recognized for many years as interesting processes to probe the
standard model (SM).  A very attractive feature of such processes is
the enhancement of the cross section and perhaps, more intriguingly, the
fact that neutrinos can scatter coherently not only on the nucleons
inside nucleus but also on the atom itself, including electrons. 

The coherent scattering takes place when momentum transfer, $q$, is
small compared with the inverse nucleus or atom size, $q R \leq
1$. For most nuclei the typical inverse sizes are in the range from 25 to
150~MeV. Therefore the condition for full coherence in the
neutrino-nuclei scattering is well satisfied for reactor neutrinos and
also for solar, supernovae neutrinos and artificial neutrino sources.

Currently, TEXONO collaboration has been considering the possibility
of detecting for the first time the coherent neutrino nucleus
scattering~\cite{TEXONO0402}. They have also pointed out that the
detection of this process will be helpful to improve the limits on the
neutrino magnetic moment. The NOSTOS collaboration is also attempting to
detect this scattering and they have plans to use this reaction for
supernovae neutrino detection~\cite{Aune:2005is}.

In this work we will show that coherent neutrino nuclei scattering is
very sensitive to other types of physics beyond the standard model
such as non-standard neutrino interactions (NSI) or additional neutral
gauge bosons, and may give constraints that are competitive with
future neutrino factories. At present, NSI involving neutrinos are
allowed to be big~\cite{Rossi1,Rossi2,Davidson:2003ha}. Moreover, NSI
have been studied in the context of solar neutrino
data~\cite{Friedland:2004pp,Guzzo:2004ue} where it has been recently
found that a large value of non-universal NSI for the neutrino
interaction with the $d$-type quark is not only allowed but also leads
to new solutions to the solar neutrino data~\cite{Miranda:2004nb}. In
the atmospheric case, although a two families
analysis~\cite{Fornengo:2001pm} gives strong constraints, the three
family case~\cite{Friedland:2004ah} showed that NSI interactions can
be big. 

Coherent neutrino nuclei processes are playing an important role in
astrophysical environments like supernovae and neutron
stars~\cite{Freedman:1977xn}. Moreover, it was recently discussed that
the inclusion of non-standard neutrino interactions may alter
supernovae explosion~\cite{Amanik:2004vm}.

A new experimental proposal able to test large NSI parameters may
forbid new solutions to the solar neutrino data and constraint these
parameters that naturally appear in many extensions of the Standard
Model. It is therefore of great interest to develop such
experiments. We will show in this article that indeed this measurement
may put strong constraints to $d$ type NSI. We will also go one step
further in this analysis to show that it would be possible to
constrain simultaneously $u$ and $d$ type NSI if we consider two
different coherent scattering experiments with two different targets
such as germanium and silicon. 

The paper is organized as follows. In the Section~\ref{sec-nsi} we
present the coherent neutrino-nuclei cross sections including
non-standard interaction contributions. In the
Section~\ref{sec-sensitivity} we estimate the expected sensitivity of
the proposed experiments and we present possible ways to improve the
sensitivity. Discussion and summary are given in the last
Section~\ref{sec-summary}.

\section{Non-standard contributions to neutrino-nucleus coherent scattering}
\label{sec-nsi}

NSI appears in many extensions of the Standard Model. For instance, in
$\nu_e d$ scattering we can have additional non-universal and flavor
changing contributions to the cross section if we consider models with
broken R parity, in which case we can have a $\nu_\tau$ in the final
state due to a squark exchange~\cite{Barger:1991ae} .  Actually, this
is just an example, but the violation of the GIM mechanism arises
naturally when considering massive neutrinos~\cite{Schechter:1980gr}
and NSI appears also in the context of the solar neutrino
analysis~\cite{NSI:solar}, Supersymmetry~\cite{NSI:susy}, models of
neutrino mass~\cite{NSI:models} etc.

In this article we parametrize the NSI contribution by using a typical
phenomenological description so that, at the four-fermion
approximation (energies $\ll M_Z$) the neutrino NSI with $u$
and $d$ quarks can be described by the effective Lagrangian
\begin{equation}
\label{NSIlagrangian}
{\cal L}_{\nu Hadron}^{NSI}=-\frac{G_F}{\sqrt{2}}
\sum_{{q=u,d}\atop{\alpha,\beta=e,\mu,\tau}}
\left[\bar\nu_\alpha \gamma^\mu (1-\gamma^5)\nu_\beta\right]
\left(
\varepsilon_{\alpha\beta}^{qL}\left[\bar q\gamma_\mu (1-\gamma^5) q\right] +
\varepsilon_{\alpha\beta}^{qR}\left[\bar q\gamma_\mu (1+\gamma^5) q\right] 
\right),
\end{equation}
where the parameters $\varepsilon_{\alpha\beta}^{qP}$ ($q=u,d$ and
$P=L,R$) describe non-standard neutrino interactions, both for the
non-universal (NU) terms, $\varepsilon_{\alpha\alpha}^{qP}$, as well
as for the flavor-changing (FC) contributions,
$\varepsilon_{\alpha\beta}^{qP}$ ($\alpha\neq\beta$).
After adding the SM Lagrangian we will get the expression 
\begin{equation}
\label{lagrangian}
{\cal L}_{\nu Hadron}^{NC}=-\frac{G_F}{\sqrt{2}}
\sum_{{q=u,d}\atop{\alpha,\beta=e,\mu,\tau}}
\left[\bar\nu_\alpha \gamma^\mu (1-\gamma^5)\nu_\beta\right]
\left(
f_{\alpha\beta}^{qL}\left[\bar q\gamma_\mu (1-\gamma^5) q\right] +
f_{\alpha\beta}^{qR}\left[\bar q\gamma_\mu (1+\gamma^5) q\right] 
\right),
\end{equation}
where the NSI parameters have been added to the SM effective coupling
constants:
\begin{eqnarray}
&&
f_{\alpha\alpha}^{uL}=\rho_{\nu N}^{NC}
\left(
\frac12-\frac23 \hat\kappa_{\nu N} \hat s_Z^2
\right) + \lambda^{uL}+ \varepsilon_{\alpha\alpha}^{uL},\nonumber\\
&&
f_{\alpha\alpha}^{dL}=\rho_{\nu N}^{NC}
\left(
-\frac12+\frac13 \hat\kappa_{\nu N} \hat s_Z^2
\right) + \lambda^{dL}+ \varepsilon_{\alpha\alpha}^{dL},\nonumber\\
&&
f_{\alpha\alpha}^{uR}=\rho_{\nu N}^{NC}
\left(
-\frac23 \hat\kappa_{\nu N} \hat s_Z^2
\right) + \lambda^{uR}+ \varepsilon_{\alpha\alpha}^{uR},
\label{couplings}\\
&&
f_{\alpha\alpha}^{dR}=\rho_{\nu N}^{NC}
\left(
\frac13 \hat\kappa_{\nu N} \hat s_Z^2
\right) + \lambda^{dR}+ \varepsilon_{\alpha\alpha}^{dR}\nonumber
\end{eqnarray}
and 
\begin{equation}
f_{\alpha\beta}^{qP} = \varepsilon_{\alpha\beta}^{qP}\,. 
\quad(\alpha\neq\beta;\;q=u,d;\;P=L,R).\nonumber
\end{equation}
Here $\hat s_Z^2=\sin^2\theta_W=0.23120$, $\rho_{\nu N}^{NC}=1.0086$,
$\hat\kappa_{\nu N}=0.9978$, $\lambda^{uL}=-0.0031$,
$\lambda^{dL}=-0.0025$ and $\lambda^{dR}=2\lambda^{uR}=7.5\times10^{-5}$
are the radiative corrections given by the 
PDG~\cite{Eidelman:2004wy}.
In the case of absence of NSI, $\varepsilon_{\alpha\beta}^{qP}=0$, 
the Lagrangian~(\ref{lagrangian}) coincides with the SM one.

From the Lagrangian in Eq.~(\ref{lagrangian}) we can obtain the 
neutrino-nucleus differential cross section which is given
by
\begin{eqnarray}
\frac{d\sigma}{dT}&=&\frac{G_F^2 M}{2\pi}\left\{
(G_V+G_A)^2+\left(G_V-G_A\right)^2\left(1-\frac{T}{E_\nu}\right)^2-
\left(G_V^2-G_A^2\right)\frac{MT}{E_\nu^2}
\right\}\,,\label{diff:cross:sect}
\end{eqnarray}
where $M$ is the mass of the nucleus, $T$ is the recoil nucleus energy, which
varies from 0 to $T_{max}=2E_\nu^2/(M+2E_\nu)$, $E_\nu$ is the
incident neutrino energy and 
\begin{eqnarray}
\label{GV}
G_V&=& 
\left[\left(g_V^p+2\varepsilon_{ee}^{uV}+\varepsilon_{ee}^{dV}\right)Z+
\left(g_V^n+\varepsilon_{ee}^{uV}+2\varepsilon_{ee}^{dV}\right)N\right]
F_{nucl}^V(Q^2)\,,\\
G_A&=& 
\left[\left(g_A^p+2\varepsilon_{ee}^{uA}+\varepsilon_{ee}^{dA}\right)
\left(Z_+-Z_-\right)+
\left(g_A^n+\varepsilon_{ee}^{uA}+2\varepsilon_{ee}^{dA}\right)
\left(N_+-N_-\right)\right]
F_{nucl}^A(Q^2)\,.
\label{GA}
\end{eqnarray}
$Z$ and $N$ represent the number of protons and neutrons in the
nucleus, while $Z_{\pm}$ ($N_\pm$) stands for the number of protons
(neutrons) with spin up and spin down respectively.
The vector and axial nuclear form factors, $F_{nucl}^V(Q^2)$ and
$F_{nucl}^A(Q^2)$, are usually assumed to be equal and of order of
unity in the limit of small energies, $Q^2\ll M^2$.

The SM neutral current vector couplings of neutrinos with
protons, $g_V^p$, and with neutrons, $g_V^n$, are defined as
\begin{eqnarray}
&&g_V^p=\rho_{\nu N}^{NC}\left(
\frac12-2\hat\kappa_{\nu N}\hat s_Z^2
\right)+
2\lambda^{uL}+2\lambda^{uR}+\lambda^{dL}+\lambda^{dR},\nonumber\\
&&g_V^n=-\frac12\rho_{\nu N}^{NC}+
\lambda^{uL}+\lambda^{uR}+2\lambda^{dL}+2\lambda^{dR}\,.
\label{vcouplings}
\end{eqnarray}
The axial
couplings can be obtained in an analogous way. 

In Eq. (\ref{diff:cross:sect}) we have included explicitly only the
non-universal NSI terms; the computation of the cross section for
flavor-changing case is straightforward.

From Eq. (\ref{diff:cross:sect}) we conclude that we can neglect axial
contributions in coherent neutrino-nuclei interactions since for most
of the nuclei the ratio of axial to vector contributions is expected
to be of order of $1/A$~\footnote{There are some cases were the axial
terms may have some importance, such as in the scattering off hydrogen
or nuclei with high spin~\cite{Gaponov:1977gr}.}.
Therefore such types of experiments will be sensitive mainly to vector
couplings, $\varepsilon_{\alpha\beta}^{qV}=
\varepsilon_{\alpha\beta}^{qL}+ \varepsilon_{\alpha\beta}^{qR}$, but
not to axial ones, $\varepsilon_{\alpha\beta}^{qA}=
\varepsilon_{\alpha\beta}^{qL}- \varepsilon_{\alpha\beta}^{qR}$, which
is similar to the case of solar neutrinos.

An explicit cancellation of all axial terms will give a better
sensitivity to vector couplings. Therefore, we will focus our study on
nuclei with even numbers of protons and neutrons with vanishing total
spin, and we will assume that the total spins of protons and neutrons,
separately, are also zero.

After some algebra and applying the assumptions already discussed, the
cross section of electron anti-neutrino coherent scattering off
a nucleus in the low energy limit, $T\ll E_\nu$, is given by
\begin{eqnarray}
\frac{d\sigma}{dT}(E_\nu,T)&=&\frac{G_F^2 M}{\pi}
\left(1-\frac{M T}{2E_\nu^2}\right)
\times\nonumber\\
&\times&\left\{\left[
Z (g_V^p+2\varepsilon_{ee}^{uV}+\varepsilon_{ee}^{dV})+
N (g_V^{n}+\varepsilon_{ee}^{uV}+2\varepsilon_{ee}^{dV})
\right]^2+\right.\nonumber\\
&+&\left.\sum_{\alpha=\mu,\tau}\left[
Z(2\varepsilon_{\alpha e}^{uV}+\varepsilon_{\alpha e}^{dV})+
N(\varepsilon_{\alpha e}^{uV}+2\varepsilon_{\alpha e}^{dV})
\right]^2 
\right\}.\label{CS}
\end{eqnarray}
In the next section we will use this expression to compute the expected number 
of events for different detectors. 

\section{Experimental sensitivity to NSI couplings}
\label{sec-sensitivity}

Experimental detection of coherent neutrino scattering has not yet been 
achieved.  Several possible methods of detecting
neutrino-nucleus coherent scattering have  been previously discussed using
superconducting~\cite{Drukier:1983gj}, acoustic~\cite{Krauss:1991ba},
cryogenic~\cite{Oberauer:2002kk}, and gas
detectors~\cite{Barbeau:2002fg,Aune:2005is}. Recently, an ultra low
energy germanium detector has been proposed by the TEXONO
Collaboration~\cite{TEXONO0402}. They plan to achieve a sub-keV
threshold with a kg-scale detector. 

In this section we demonstrate the sensitivity to NSI coming from the
coherent neutrino-nuclei scattering. In order to apply our analysis to
a concrete case, we will concentrate our discussion on the germanium
TEXONO proposal~\cite{TEXONO0402}. The detector would be located
at the Kuo-Sheng Nuclear Power Station at a distance of 28 m from the
reactor core. We asume a typical neutrino flux of
$10^{13}$~s$^{-1}$~cm$^{-2}$. There are several parametrizations that
consider in detail the neutrino spectrum coming from a
reactor~\cite{Huber:2004xh}, depending on the fuel composition. In
this work we prefer to consider only  the main
component of the spectrum~\cite{Schreckenbach:1985ep} coming from
$^{235}$U, since the experiment is not running yet. For energies
below $2$~MeV there are only theoretical calculations for the
antineutrino spectrum that we take from Ref.~\cite{Kopeikin:1997ve}.

Besides the TEXONO proposal, we will also discuss the more theoretical
case of a silicon detector in order to illustrate the potential of a combined
analysis of two different materials.

\subsection{The detector}

If we neglect for a moment the detector efficiency and resolution, we
can estimate the total number of expected events in a detector as
\begin{equation}
N_{\rm{events}}=t\phi_0\frac{M_{\rm{detector}}}{M}
\int\limits_{E_{min}}^{E_{max}}dE_\nu
\int\limits_{T_{th}}^{T_{max}(E_\nu)}dT
\lambda(E_\nu)\frac{d\sigma}{dT}(E_\nu,T)\,,
\label{Nevents}
\end{equation}
with $t$ the data taking time period, $\phi_0$ the total neutrino
flux, $M_{\rm{detector}}$ the total mass of the detector,
$\lambda(E_\nu)$ the normalized neutrino spectrum, $E_{max}$ the
maximum neutrino energy, $T_{th}$ the detector energy threshold. One
can see that the minimum required incoming neutrino energy 
depends on the detector threshold $T_{th}$ and the nucleus mass $M$
through the relation $T_{th}(E_\nu)=2E_\nu^2/(M+2E_\nu)$. For
instance, for a $400$~eV detector's threshold and a $^{76}$~Ge
nucleus, the minimum required incoming neutrino energy is
$E_{min}=3.8$~MeV, which is well satisfied by reactor neutrinos.

One can see from Eq.~(\ref{CS}) that, since the number of NSI
parameters is big, there are cancellations among them and therefore a
degeneracy in their determination, both for non-universal and for
flavor changing cases. 

Although in many cases it is convenient to reduce the number of
parameters (by neglecting the $u$ type NSI, for example), we will
demonstrate that it is possible to minimize the degeneracy by choosing
particular nuclei. We will introduce in the following a simple
criteria to find such detectors.

Let's consider first the case of non-universal NSI parameters,
$\varepsilon_{ee}^{qV}$, assuming that flavor changing parameters are
negligible, $\varepsilon_{\alpha e}^{qV}=0$, $\alpha\neq e$. One can
easily see that $\varepsilon_{ee}^{uV}$ and $\varepsilon_{ee}^{dV}$
are related as
\begin{equation}
\varepsilon_{ee}^{uV}(A+Z)+\varepsilon_{ee}^{dV}(A+N)=\mbox{const}\,.
\label{degk}
\end{equation}
The slope of the curve on the plane ($\varepsilon_{ee}^{uV}$,
$\varepsilon_{ee}^{dV}$) is described by the ratio $k=(A+N)/(A+Z)$.
It is clear therefore that, in order to constrain both parameters
efficiently, two criteria should be taken into account:
\begin{itemize}
\item one should choose at least two different nuclei, for which the
parameter $k$ is maximally different;
\item one should choose nuclei for which the number of
expected events is larger.
\end{itemize}
In contrast to naive expectations the second criteria is non-trivially
related with the mass of nucleus, since the maximum recoil energy,
$T_{max}$, also depends on nucleus mass. 

These arguments are illustrated in Fig.~\ref{fig:bestelements} where
we consider the expected number of events for a one kg detector, after
one year of data taking, with either 100 eV or 400 eV threshold. In
this figure we show iso-curves of equal number of events depending on
the number of protons and neutrons of the nuclei. We show in the same
plot some nuclei that have been considered as possible detectors in
the literature. Also, as an illustration, lines for $k=1$ and $k=1.15$
are shown. From this plot it is easy to see that germanium is
actually a good material for this kind of experiments since it lies in
a region where the expected number of events is high and the $k$
parameter is big. We can also see that silicon could have a comparable
number of events and the smallest possible value $k=1$. Although at
present there is no experimental proposal to use such a material, we
will also consider it in our analysis since the theoretical
expectations of having complementary information from it are very
tempting.

Although in the following sections we will discuss both the case of
$^{28}$Si ($k=1$) as well as $^{76}$Ge
($k\approx1.11$)~\cite{TEXONO0402}, it is also important to note that
the computations in Fig.~\ref{fig:bestelements} were done for a fixed
detector mass. Therefore, although in our plot the expected number
of events for  Xenon is lower than germanium for the energy threshold
$T_{th}=400$~eV, this can be compensated by a larger amount of
material, as seems to be the case of the corresponding NOSTOS
experiment proposal~\cite{Aune:2005is}.

\FIGURE{
\includegraphics[width=0.45\textwidth]{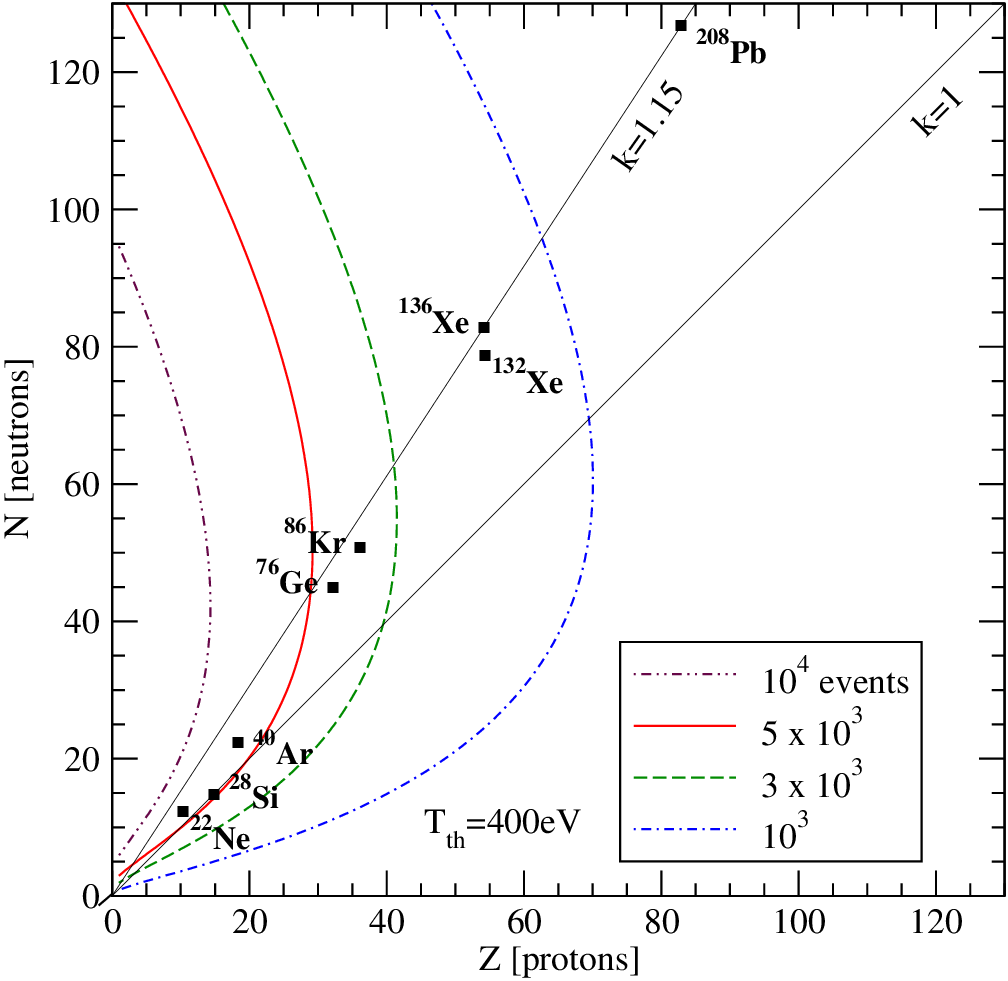}%
\hspace{0.01\textwidth}
\includegraphics[width=0.45\textwidth]{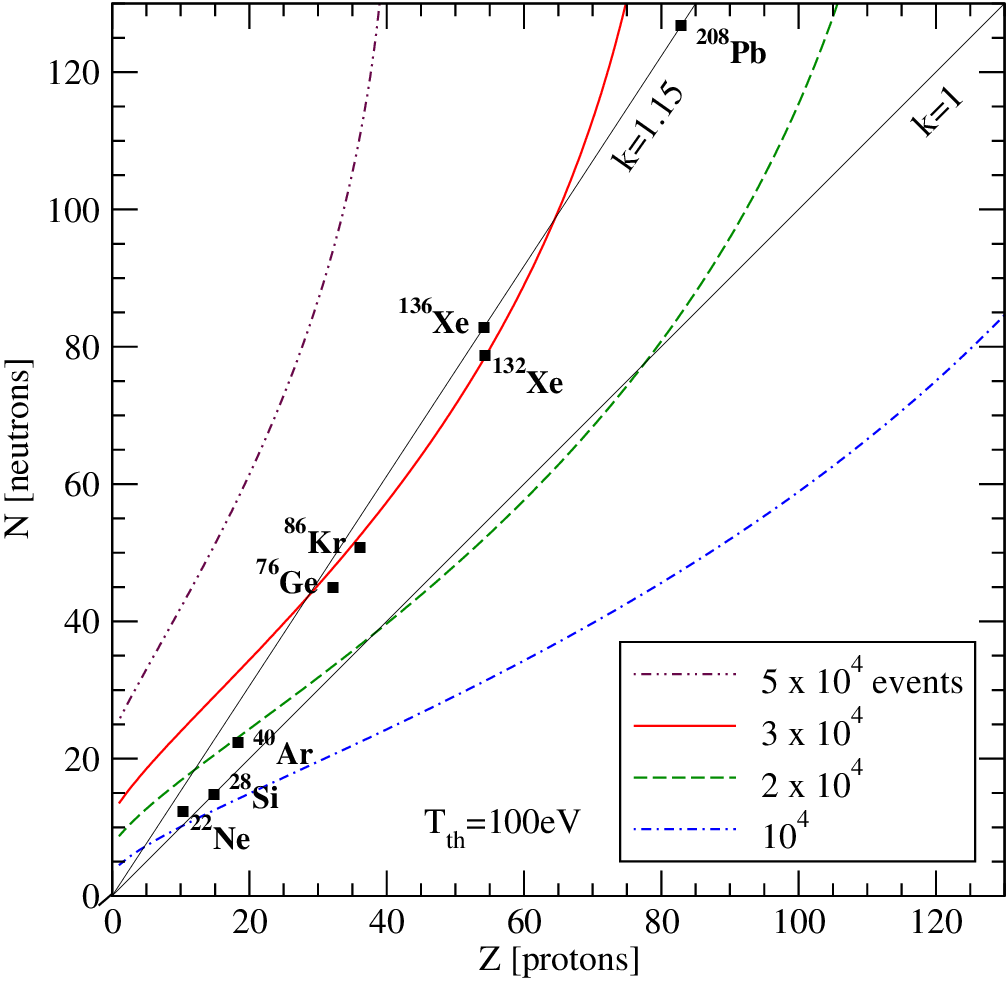}%
\caption{\label{fig:bestelements} Iso-curves of the expected number of
events per kg per year are shown on the plane N
{\it vs} Z for two values of energy threshold,
$T_{th}=400$~eV (left panel) and $100$~eV (right panel). NSI contribution is
neglected.  Different elements are pointed. Two solid lines are shown
for two values of parameter $k=1$ and $1.15$.}
}

\subsection{Non-universal NSI}

We start our NSI analysis by assuming only non-universal parameters,
and neglecting the flavor changing ones. We will see that in
this case, for constraining both $u$ type and $d$ type NSI at a time
it is necessary to perform two different experiments. However, if we
are interested in constraining only $d$ type NSI a germanium
experiment can get a constraint competitive with future neutrino
factories.

The expected number of events for  germanium and silicon 
were calculated assuming a detector mass of $1$~kg, $1$~year of data
taking, and two possible experimental threshold energies were
considered: $T_{th}=100$eV and $T_{th}=400$eV.  To calculate the
sensitivity we assumed that both experiments will measure exactly the SM
expected number of events and we perform a $\chi^2$ analysis based
only on statistical errors. 

The results are shown in Fig.~\ref{fig:gesi} both for the case of a
germanium detector (uncolored lines) as well as for the combined analysis 
of germanium plus silicon. It is possible to see that for the case of 
one experiment, both $u$ type and $d$ type NSI parameters are strongly 
correlated while the combination of two experiments can give a simultaneous 
determination of both parameters at the order of few percent. 

In order to compare the experimental reach with other future
experiments, note that the expected sensitivity for NSI
parameter $\varepsilon^{dV}_{ee}$ in neutrino factories, when all
other parameters vanish, is approximately
0.006~\cite{Davidson:2003ha} while the expected sensitivity for the
germanium only case for a $400$ eV threshold is even better
(0.003). We can also note that in the case of considering only one
parameter at a time, the sensitivity doesn't improve if we consider
the case of a silicon detector. However, such an additional experiment
could be very helpful in constraining both $\varepsilon^{dV}_{ee}$ and
$\varepsilon^{uV}_{ee}$ at the same time, especially for the case of
$100$~eV energy threshold.

\FIGURE{
\includegraphics[width=0.45\textwidth]{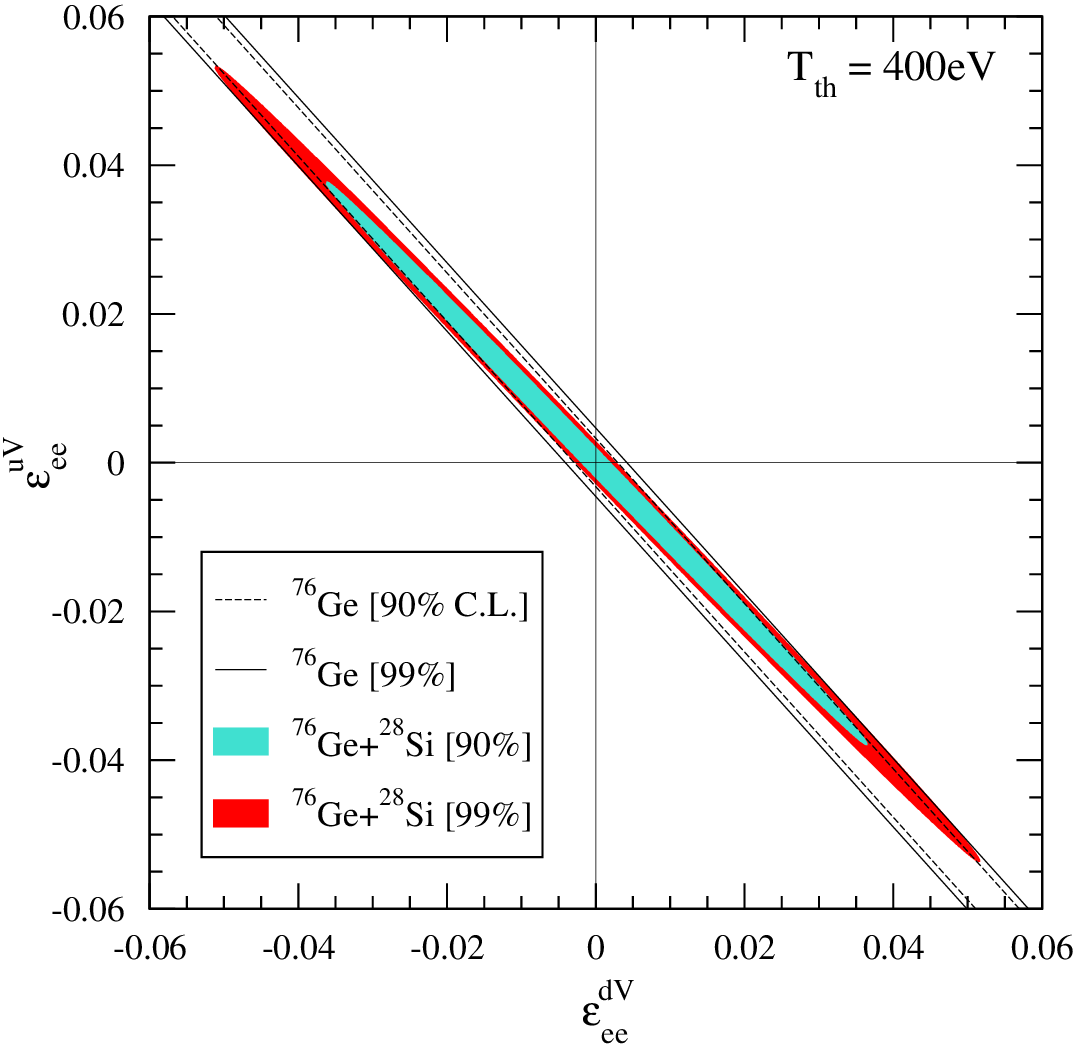}%
\hspace{0.01\textwidth}
\includegraphics[width=0.45\textwidth]{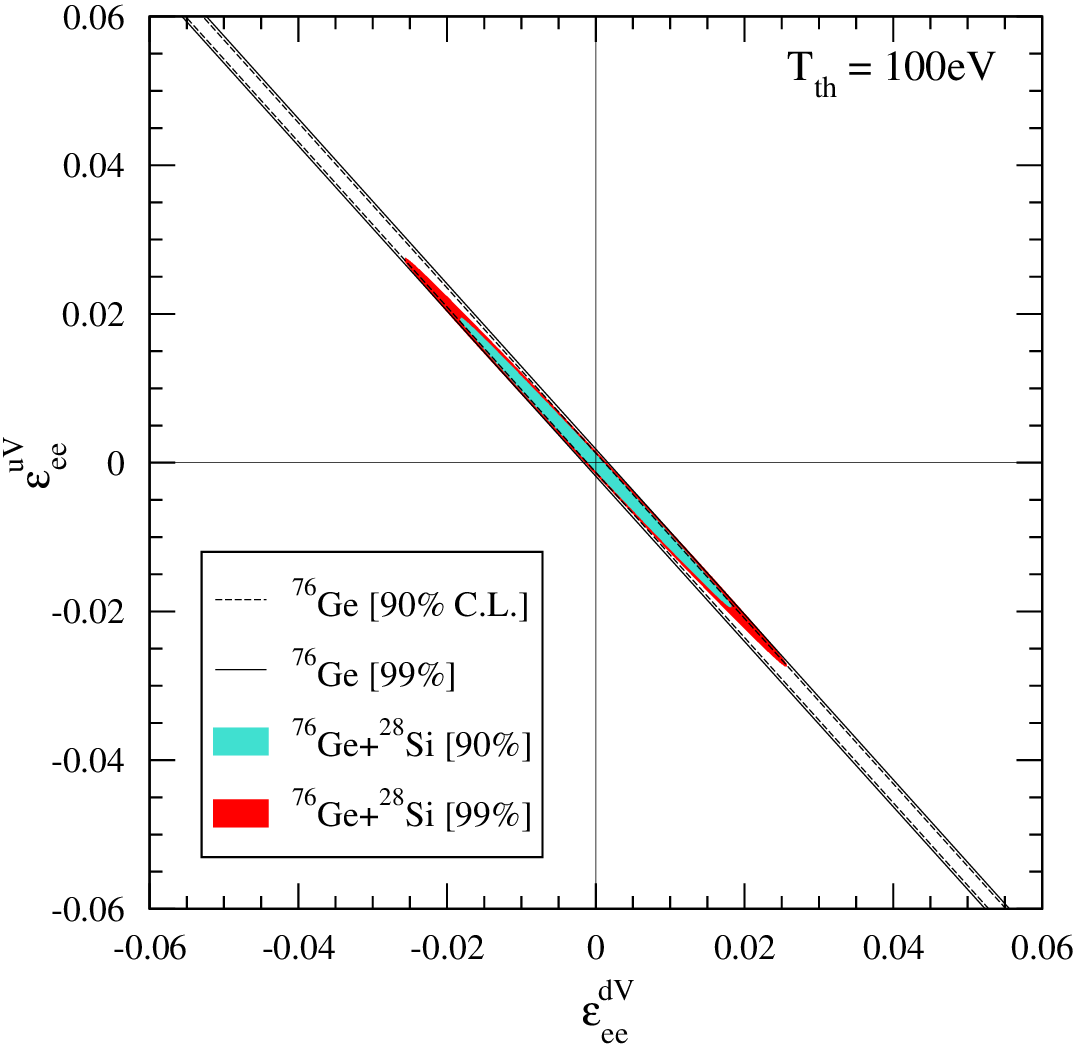}%
\caption{\label{fig:gesi}The allowed regions of non-universal NSI
parameters $\varepsilon_{ee}^{uV}$ and $\varepsilon_{ee}^{dV}$ are
shown at 90 and 99\%~C.L. for combined data from two detectors of
$^{28}$Si and $^{76}$Ge (colored regions) and only for $^{76}$Ge
(solid and dashed lines). It was assumed 1~kg mass and 1~year of data
taking for both detectors, only statistical errors are taken into
account. The results presented for two values of threshold,
$T_{th}=400$~eV (left) and $100$~eV (right panel).}
}

\subsection{Flavor-changing NSI}

Now we consider the case of flavor changing NSI neglecting
non-universal terms. For this analysis we also take into account the
strong bounds on flavor changing parameters for muon neutrinos,
$\varepsilon_{\mu e}^{qV}<7.7\times10^{-4}$, derived from $\mu-e$
conversion on nuclei~\cite{Davidson:2003ha}, which motivates us to
perform our analysis only for flavor changing NSI parameters that
includes tau-neutrinos in the final state, $\varepsilon_{\tau e}^{uV}$
and $\varepsilon_{\tau e}^{dV}$.

The degeneracy problem discussed in the previous section for
non-universal interactions is analogous in the flavor changing NSI
case and it is described by the same parameter $k$, defined after
Eq.~(\ref{degk}). Because of that we consider the same materials for
detectors and we make the same assumptions. We give our results for
these parameters in Fig.~\ref{fig:gesi.fc}. In this case the expected
sensitivity ($\varepsilon_{\tau e}^{dV}\simeq 0.03$.) is again better
than that expected for neutrino factories~\cite{Davidson:2003ha}
$\varepsilon_{\tau e}^{dV}\simeq 0.06$.
\FIGURE{
\includegraphics[width=0.45\textwidth]{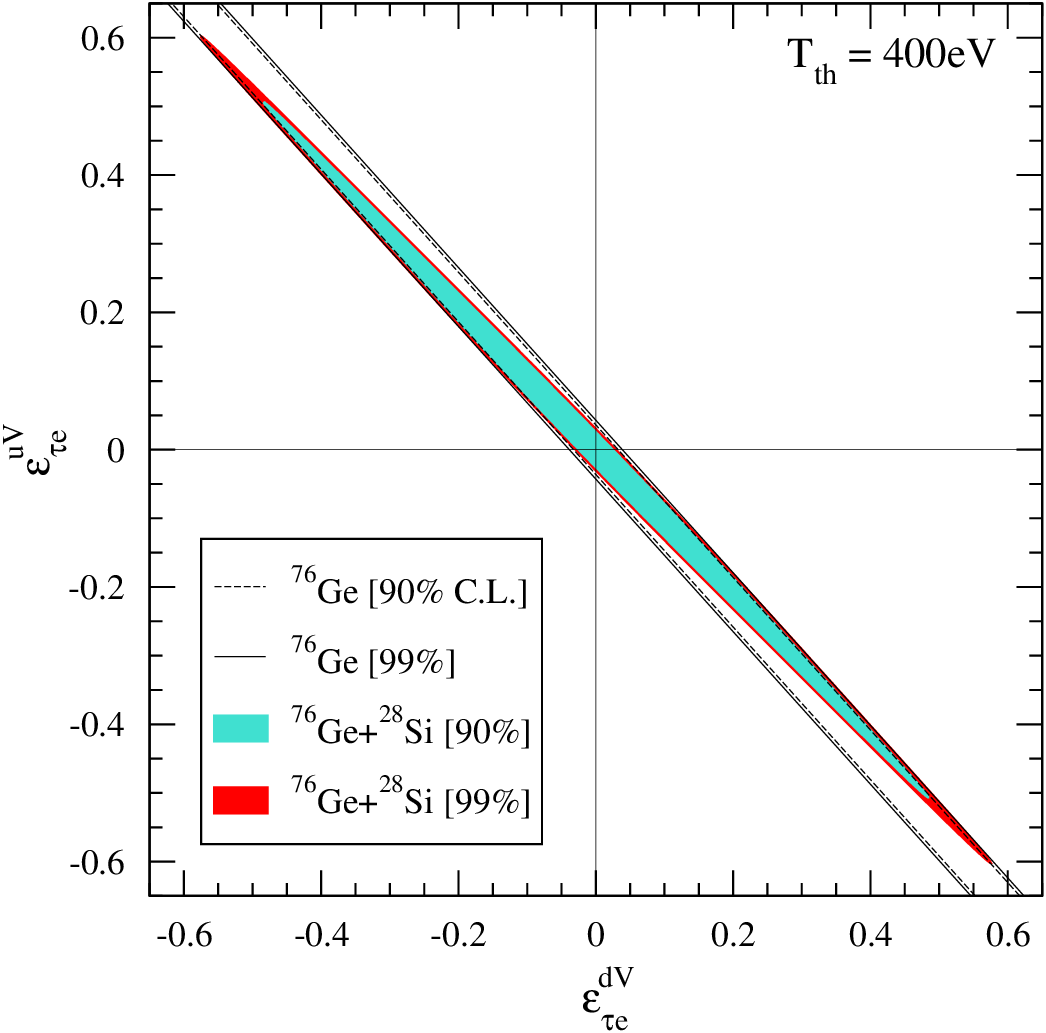}%
\hspace{0.01\textwidth}
\includegraphics[width=0.45\textwidth]{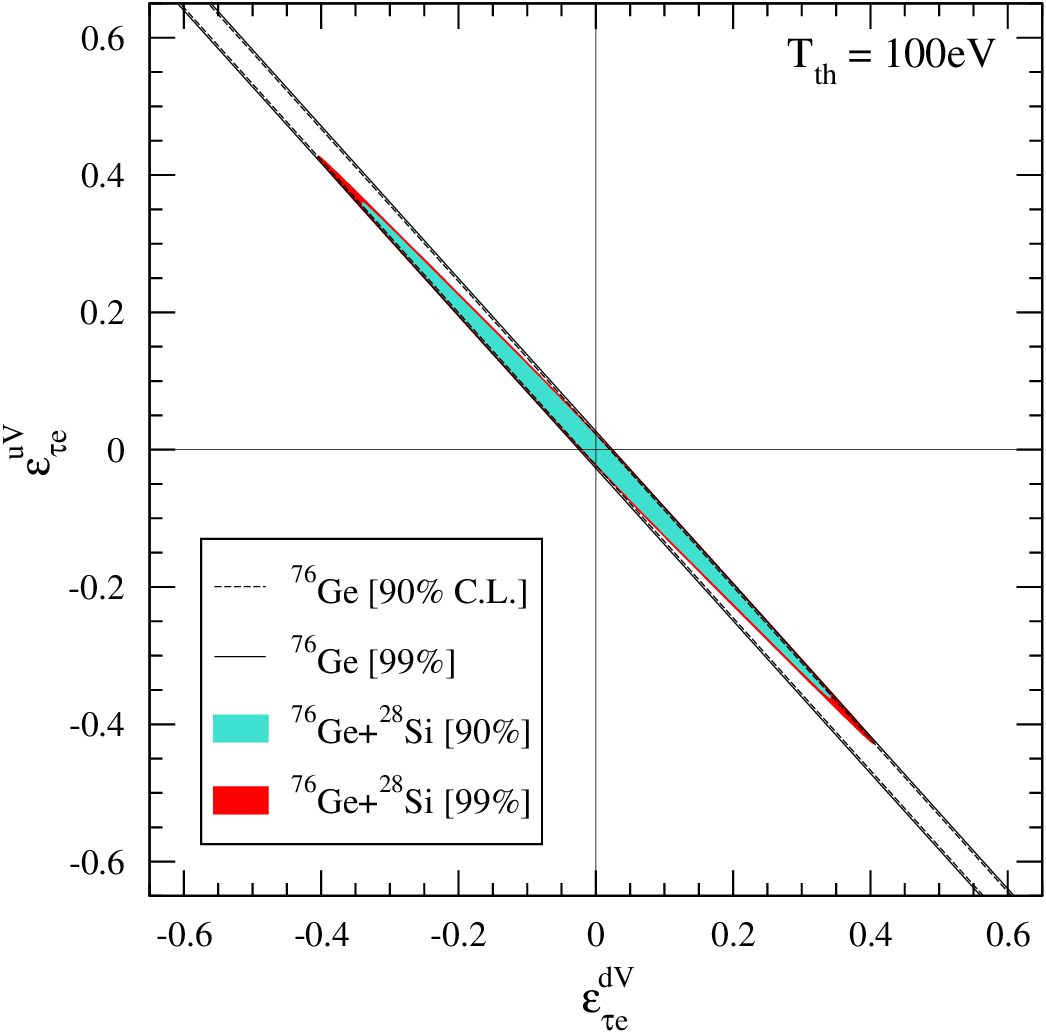}%
\caption{\label{fig:gesi.fc}The same as Fig.~\ref{fig:gesi} but the allowed
regions of flavor-changing NSI parameters $\varepsilon_{\tau e}^{uV}$
and $\varepsilon_{\tau e}^{dV}$ are shown.}  
}

\subsection{Non-universal {\it vs} flavor-changing NSI for $d$-quark}

Here we compute the potential of this proposal when both non-universal
and flavor changing NSI are considered one at a time, with NSI with
$d$-type quark only, neglecting NSI with $u$-quark. Besides
theoretical motivations, it is also interesting to study this case
because, as it has been pointed out in the introduction, for this case
is possible to have additional solutions to the solar neutrino
problem, if the non-universal NSI contribution is
big~\cite{Miranda:2004nb}.  As before we consider only NSI parameters
which include tau-neutrinos, since the NSI muon neutrino interaction is
strongly constrained.

As one can see from Eq.~(\ref{CS}), in this case we have again a
degeneracy in the determination of the parameters $\varepsilon_{e
e}^{dV}$ and $\varepsilon_{\tau e}^{dV}$. The allowed region on the
plane ($\varepsilon_{e e}^{dV}$, $\varepsilon_{\tau e}^{dV}$) has the
form of an elliptic band. The width of the ellipse is twice as big as
the height, which is described by the parameter,
$a=(Zg_V^p+Ng_V^n)/(Z+2N)$, the analog of the parameter $k$ used in
previous cases.

In the same way as it was discussed in the previous sections, we can
use two different elements to break this degeneracy using as a guidance
the parameter $a$. It is straightforward to check that the same
materials discussed above are also good for the present
case. By doing a similar analysis as in previous subsections we
obtained, for the combined detectors case, an allowed region
which is simply connected.  These results are shown in
Fig.~\ref{fig:gesi.nu.fc} from which it is clear that the combination 
of the two studied experiments could definitely probe additional solutions 
to the solar neutrino data. 
\FIGURE{
\includegraphics[width=0.45\textwidth]{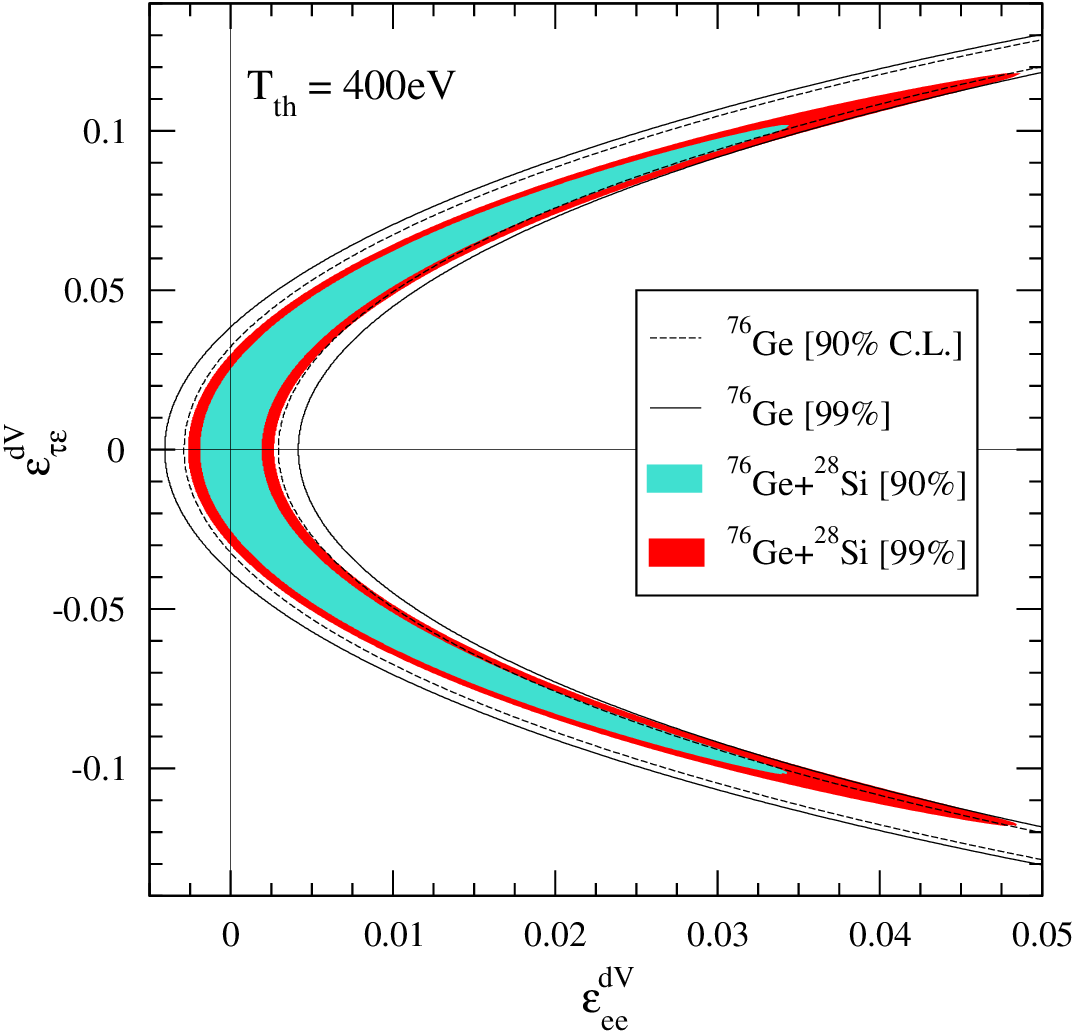}%
\hspace{0.01\textwidth}
\includegraphics[width=0.45\textwidth]{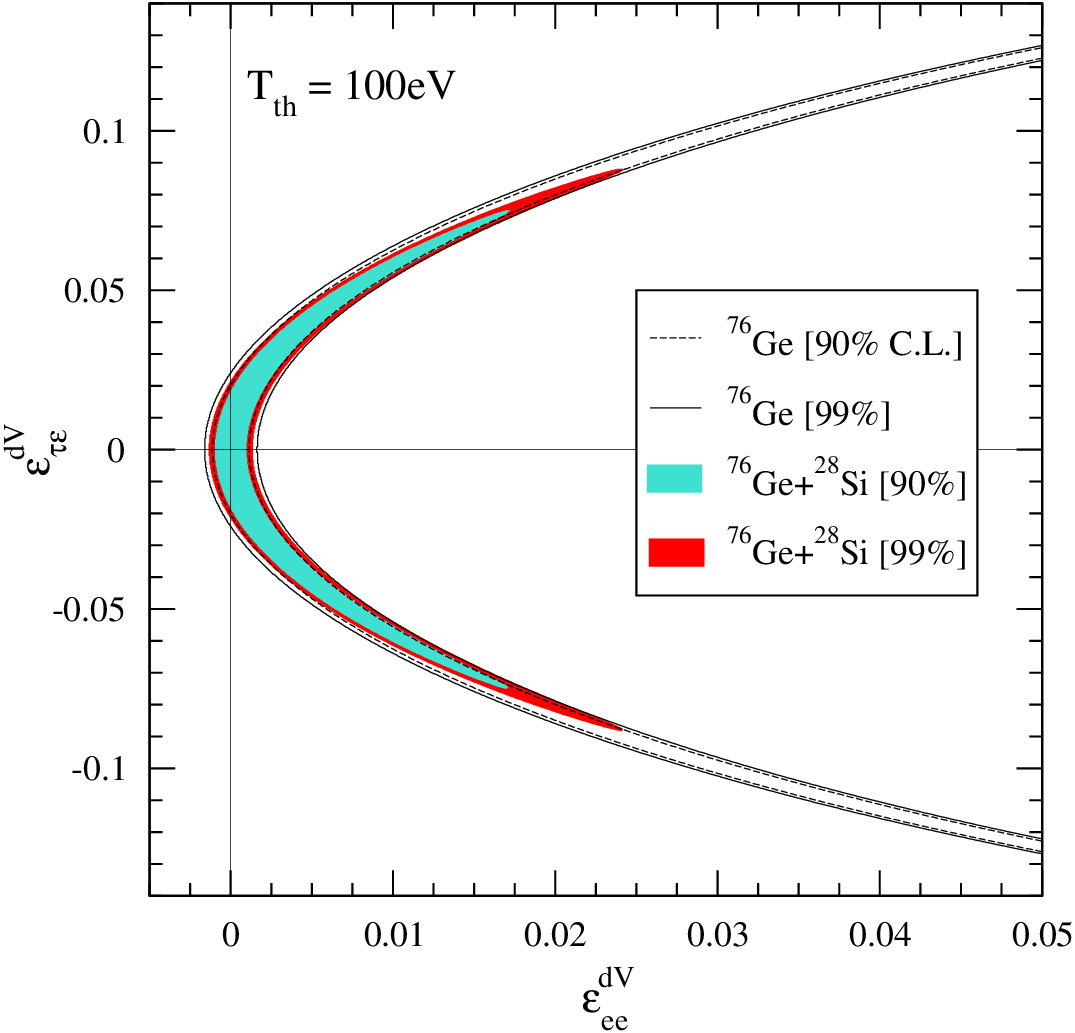}%
\caption{\label{fig:gesi.nu.fc}The same as Fig.~\ref{fig:gesi} but the
allowed regions of non-universal {\it vs} flavor-changing NSI
parameters $\varepsilon_{e e}^{dV}$ and $\varepsilon_{\tau e}^{dV}$
for $d$ type quark only are shown.}
}

\subsection{Extra heavy neutral gauge boson $Z'$}

\TABLE[!t]  {
    \caption{Quantum numbers for the light particles in the {\bf 27} of $E_6$.}
    \label{tab:E6}
        \begin{tabular}{cccc}
            \hline {\rule[-3mm]{0mm}{8mm}} & $T_3$ & $\sqrt{40}Y_\chi$ &
            $\sqrt{24}Y_\psi$   \\
            \hline
            $Q$   & $\pmatrix{1/2 \\ -1/2}$ & $-1$ & $1$ \\
            $u^c$ &       $0$               & $-1$ & $1$ \\
            $e^c$ &       $0$               & $-1$ & $1$ \\
            $d^c$ &       $0$               & $ 3$ & $1$ \\
            $l$   & $\pmatrix{1/2 \\ -1/2}$ & $ 3$ & $1$ \\
            \hline
        \end{tabular}  
}
As we have discussed above, NSI appears in many extensions of the SM.
As an illustration of the coherent neutrino-nuclei scattering we would
like to discuss a specific example; we consider in this subsection
the particular case of an additional neutral gauge boson $Z'$ that
arises from a primordial $E_6$ gauge
symmetry~\cite{Gonzalez-Garcia:1990yq}.  The new boson $Z'$ affects
the neutral current couplings of the SM since these extension usually
involve an extra $U(1)$ hypercharge symmetry at low energies that may
be given as the mixture of those associated with the symmetries
$U(1)_\chi$ and $U(1)_\psi$. We show the quantum numbers for the SM particles 
in Table \ref{tab:E6}.

The corresponding hyper-charge is then specified by
\begin{equation}
Y_\beta =\mbox{cos}\beta Y_{\chi}+\mbox{sin} \beta Y_{\psi}, 
\end{equation}
while the charge operator is given as $ Q=T^3+ Y $.  Any value of
$\beta$ is allowed, giving us a continuum spectrum of possible models
of the weak interaction.

At tree level it is possible to write an expression for the effective
four-fermion Lagrangian describing low-energy neutral current
phenomena. We neglect non-standard radiative corrections because its
contribution is of order $(\alpha/\pi)(M_Z^2/M_{Z'}^2)$~\cite{Cvetic1987}.
The prescription is exactly an NSI type Lagrangian
such as eq. (\ref{lagrangian}) with modified expression for the
electroweak coupling constants, 

\begin{eqnarray}
\label{couplingsnewz}
\varepsilon_{ee}^{uL}&=&- 4 \gamma \sin^2\theta_W \rho_{\nu N}^{NC}
\left({c_\beta \over \sqrt{24}}-{s_\beta \over 3}\sqrt{5 \over 8} \right)
\left({3 c_\beta \over 2 \sqrt{24}}+{s_\beta \over 6}\sqrt{5 \over 8} \right)
\nonumber \\
\varepsilon_{ee}^{dL}&=&\varepsilon_{ee}^{uL}
\nonumber \\
\varepsilon_{ee}^{uR}&=&-\varepsilon_{ee}^{uL}\nonumber \\
\varepsilon_{ee}^{dR}&=&-8 \gamma \sin^2\theta_W \rho_{\nu N}^{NC}
\left({3 c_\beta \over 2 \sqrt{24}}+
{s_\beta \over 6}\sqrt{5 \over 8} \right)^2 ,
\end{eqnarray}
where $c_\beta=\mbox{cos}\beta$, $s_\beta=\mbox{sin}\beta$ and
$\gamma=(M_Z/M_{Z'})^2$.  Three main models have been extensively
studied, namely: the $\chi$ model ($\mbox{cos}\beta=1$), the $\psi$ model
($\mbox{cos}\beta=0$) and the $\eta$ model (cos$\beta=\sqrt{3/8}$). In
previous articles~\cite{Miranda:1997vs} it has been stressed that low
energy neutrino experiments are more sensitive to the $\chi$ model
than to others $E_6$ models and therefore we concentrate in this
particular case.

We repeat now our analysis for NSI but with the couplings of the
$\chi$ model that depends on the mass of the extra neutral gauge boson
$M_{Z'}$. Considering only the germanium detector case we found a
potential sensitivity, at 90 \% C. L., of $M_{Z'}>930$~GeV for a
$400$~eV threshold and $M_{Z'}> 1420$~GeV for a $100$~eV threshold.
For the case of a combined germanium plus silicon detector the
expectations are even better, with a lower bound of $1160$~GeV for a
$400$~eV threshold and $1640$~GeV for the case of a $100$~eV threshold
at 90 \% C.L.

Although the case of an extra neutral gauge boson comming from $E_6$
models is just one example among many other extensions that can be
parametrized by NSI, it is very useful to illustrate the potential of
neutrino coherent scattering off nucleus since the sensitivity in this
case is quite competitive with the current Tevatron constraints and an
improvement will depend on the specific experimental set up.

\section{Discussion and summary}
\label{sec-summary}

We have discussed the physics potential of a coherent neutrino nucleus
scattering measurement to test both NSI and extra neutral gauge
bosons. 
\TABLE[!t]{
    \caption{Constraints on NSI parameters at 90\% C.L. taking one
    parameter at a time in different experiments. 
    Present limits and $\nu$-factory sensitivity are from 
   Ref.~\cite{Davidson:2003ha}.    \label{tab:1param}}
        \begin{tabular}{|c|c|c|c|c|}
            \hline      {\rule[-3mm]{0mm}{8mm}  } & Present Limits  & $\nu$ Factory &
            $^{76}$Ge $_{T_{th}=400eV}$ &  $^{76}$Ge$+^{28}$Si $_{T_{th}=400eV}$ \\
            & & & ($^{76}$Ge $_{T_{th}=100eV}$) &($^{76}$Ge$+^{28}$Si $_{T_{th}=100eV}$)
               \\\hline {\rule[-3mm]{0mm}{8mm}  }
             $\eps_{ee}^{dV}$ &$-0.5 < \eps_{ee}^{dV} < 1.2  $ &$ |\eps_{ee}^{dV}| < 0.002 $ 
            & $|\eps_{ee}^{dV}| < 0.003$  &  $ |\eps_{ee}^{dV}| < 0.002$   \\ 
               & & &($|\eps_{ee}^{dV}| < 0.001$)  &  ($ |\eps_{ee}^{dV}| < 0.001$)   \\ 
            \hline {\rule[-3mm]{0mm}{8mm}  }
             $\eps_{\tau e}^{dV}$ & $|\eps_{\tau e}^{dV}| < 0.78 $ & $| \eps_{\tau e}^{dV}| < 0.06 $ 
            & $|\eps_{\tau e}^{dV}| < 0.032 $ & $| \eps_{\tau e}^{dV}| < 0.024$   \\ 
               & & &($|\eps_{\tau e}^{dV}| < 0.020$)  &  ($ |\eps_{\tau e}^{dV}| < 0.017$)   \\ 
            \hline {\rule[-3mm]{0mm}{8mm}  }
             $\eps_{ee}^{uV}$& $ -1.0 < \eps_{ee}^{uV} < 0.61 $ & $ |\eps_{ee}^{uV}| < 0.002 $ 
            &$ |\eps_{ee}^{uV}| < 0.003$ & $| \eps_{ee}^{uV} |< 0.002$   \\ 
               & & &($|\eps_{ee}^{uV}| < 0.001$)  &  ($ |\eps_{ee}^{uV}| < 0.001$)   \\ 
            \hline {\rule[-3mm]{0mm}{8mm}  }
             $\eps_{\tau e}^{uV}$ &$ |\eps_{\tau e}^{uV}| < 0.78 $ & $ |\eps_{\tau e}^{uV}| < 0.06 $ 
            & $ |\eps_{\tau e}^{uV}| < 0.036 $& $ |\eps_{\tau e}^{uV}| < 0.023$   \\ 
               & & &($|\eps_{\tau e}^{uV}| < 0.023$)  &  ($ |\eps_{\tau e}^{uV}| < 0.018$)   \\ 
            \hline
        \end{tabular}  
}
\TABLE[!t]  {
    \caption{Constraints on NSI parameters at 90\% C.L.taking two parameters at a time.}
    \label{tab:2params}
        \begin{tabular}{|c|c|c|}
            \hline      {\rule[-3mm]{0mm}{8mm}  } & $^{76}$Ge$+^{28}$Si $_{T_{th}=400eV}$ &$^{76}$Ge$+^{28}$Si $_{T_{th}=100eV}$ \\
            \hline
             $\eps_{ee}^{dV}$ &$|\eps_{ee}^{dV}|<0.036$ &$|\eps_{ee}^{dV}|<0.018$ \\
             $\eps_{ee}^{uV}$ &$|\eps_{ee}^{uV}|<0.038$ &$|\eps_{ee}^{uV}|<0.019$ \\    
            \hline {\rule[-3mm]{0mm}{8mm}  }
             $\eps_{\tau e}^{dV}$ &$|\eps_{\tau e}^{dV}|<0.48$ &$|\eps_{\tau e}^{dV}|<0.34$ \\
             $\eps_{\tau e}^{uV}$ &$|\eps_{\tau e}^{uV}|<0.50$ &$|\eps_{\tau e}^{uV}|<0.37$ \\  
            \hline {\rule[-3mm]{0mm}{8mm}  }
             $\eps_{ee}^{dV}$ &$-0.002<\eps_{ee}^{dV}<0.034$ &$-0.0009<\eps_{ee}^{dV}<0.016$ \\
             $\eps_{\tau e}^{dV}$ &$|\eps_{\tau e}^{dV}|<0.1$ &$|\eps_{\tau e}^{dV}|<0.074$ \\  
            \hline
        \end{tabular}  
}
We have shown that such experiments could give constrains to NSI that
are competitive with those coming from a future neutrino
factory. Moreover, if we go one step further and consider two
different experiments, we can constrain simultaneously $u$ and $d$
type NSI. We illustrate the physics reach of these experiments in
Table~\ref{tab:1param}, where we compare the expected NSI sensitivity
in the case of a neutrino factory and in the case of either one or
two neutrino coherent scattering experiments. In this Table, we have
considered the sensitivity to one parameter at a time. In
Table~\ref{tab:2params} we consider two different parameters
simultaneously for the case of a combined analysis with two neutrino
coherent scattering experiments. It is possible to see that these two
experiments are very effective in constraining NSI parameters.

In summary we have shown that neutrino coherent scattering off nuclei
has a lot of potential to make precision tests of new physics, not
only in the case of a non-zero neutrino magnetic moment as had already
been discussed, but also to test non-standard neutrino interactions
and extra heavy neutral gauge bosons. 

\begin{acknowledgments}
  We kindly thank for very productive discussions to H. T. K. Wong, M.
  Fairbairn and D. Julio. This work has been supported by CONACyT,
  SNI-Mexico. TIR was supported by RFBR grant {04-02-16386-a}.  TIR
  thanks Physics Department of CINVESTAV for the hospitality during
  the visit when this work was done.
\end{acknowledgments}

\end{document}